\documentclass[a4paper,11pt]{article}
\pdfoutput=1 

\usepackage{jcappub} 
\usepackage[T1]{fontenc} 

\usepackage{graphicx,amssymb,amsmath,amsthm,amsfonts,epsfig,times,bm}
\usepackage{aas_macros}
\usepackage{cleveref}

\usepackage{amsmath,amssymb}
\usepackage{tensor}

\def\be{\begin{equation}}
\def\ee{\end{equation}}
\newcommand{\bea}{\begin{eqnarray}}
\newcommand{\eea}{\end{eqnarray}}

\newcommand{\dd}{\text{d}}
\def\l{\left}
\def\r{\right}
\def\p{\partial}

\def\w{\omega}

\title{\boldmath Axion boson stars}

\author[1]{Davide Guerra,}\emailAdd{davide.guerra3@gmail.com}
\author[2]{Caio F. B. Macedo,}\emailAdd{caiomacedo@ufpa.br}
\author[1,3]{Paolo Pani,}\emailAdd{paolo.pani@uniroma1.it}

\affiliation[1]{Dipartimento di Fisica, ``Sapienza'' Universit\`a di Roma \& Sezione INFN Roma1, Piazzale Aldo Moro 
5, 00185, Roma, Italy}
\affiliation[2]{Faculdade de F\'isica, Universidade Federal do Par\'a, Salin\'opolis, Par\'a, 68721-000 Brazil
}
\affiliation[3]{Scuola Superiore di Studi Avanzati Sapienza, Viale Regina Elena 291, 00161, Roma, Italy}

\abstract
{
We study novel solitonic solutions to Einstein-Klein-Gordon theory in the presence of a periodic scalar potential 
arising in models of axion-like particles. The potential depends on two parameters: the mass of the scalar field $m_a$
and the decay constant $f_a$; the standard case of the QCD axion is recovered when $m_a\propto1/f_a$. When 
$f_a\to\infty$ the solutions reduce to the standard case of ``mini'' boson stars supported by a massive free scalar 
field. As the energy scale $f_a$ of the scalar self-interactions decreases we unveil several novel features of the 
solution: new stability branches emerge at high density, giving rise to very compact, radially stable, boson stars. 
Some of the most compact configurations acquire a photon sphere. When $f_a$ is at the GUT scale, a boson star 
made of QCD axions can have a mass up to ten solar masses and would be more compact than a neutron star.
Gravitational-wave searches for these exotic compact objects might provide indirect evidence for ultralight axion-like 
particles in a region not excluded by the black-hole superradiant instability.
}

\begin{document}

\maketitle
\flushbottom

\section{Introduction}\label{intro}
The nature of dark matter remains one of the greatest mysteries in physics. The most compelling hypothesis is that 
dark matter is made of new particles beyond the Standard Model, but their properties mostly unknown. 
The mass of particle dark matter candidates spans a range of several tens of orders of magnitude (with candidates as 
light as $10^{-22}\,{\rm eV}$ 
or as massive as $10^{4}\,{\rm GeV}$). One of the most compelling dark-matter candidates is the 
\emph{axion}~\cite{Peccei:1977hh,PhysRevLett.40.223,PhysRevLett.40.279}, a 
pseudo Nambu-Goldstone boson which acquires a non-derivative coupling to the Standard Model only due to topological 
charges~\cite{Preskill:1982cy,Davidson:2016uok,Baer:2014eja,Carosi:2012zz,diCortona:2015ldu,Klaer:2017ond}. While 
originally introduced by Peccei and Quinn to solve the strong CP problem in quantum chromodynamics~(QCD), the axion 
later served as a prototype for weakly-interacting ultralight bosons beyond the Standard 
Model~\cite{Jaeckel:2010ni,Essig:2013lka,Goodsell:2009xc,Arvanitaki:2010sy}. For instance, axion-like particles~(ALPs) 
with various masses are ubiquitous in string-theory compactification and have been suggested as a generic signature of 
extra dimensions~\cite{Svrcek:2006yi,Arvanitaki:2009fg}. Several experiments are ongoing to search for ALPs in a 
wide mass range (for a recent review see Ref.~\cite{Irastorza:2018dyq}).

Ultralight bosons might form macroscopic Bose-Einstein condensates obtained as localized and coherently oscillating 
solutions to their classical field equations. When the boson mass is $\sim 10^{-22}\,{\rm eV}$, these condensates 
provide a natural alternative to the standard structure formation through dark matter 
seeds~\cite{Guzman:2009zz,Suarez:2013iw,Li:2013nal,Hui:2016ltb,Dev:2016hxv,Chavanis:2016dab}. When the boson mass is heavier, these condensates are 
smaller and can have the typical size and mass of an astrophysical compact object. These solutions are generically known 
as \emph{boson stars}~(BSs) (when the field is complex) or \emph{oscillatons}~\cite{Seidel:1991zh} (when the field is 
real), and provide the prototypical example of an exotic compact object~(ECO)~\cite{Giudice:2016zpa}, namely a 
hypothetical dark compact star that is neither a black hole nor a neutron star~\cite{Cardoso:2017cqb,Cardoso:2019rvt}.

More specifically, a BS (also known as Klein-Gordon geon) is a stationary configuration of a
scalar field bounded by gravity~\cite{Kaup:1968zz,Ruffini:1969qy,Khlopov:1985jw,Seidel:1991zh} (see 
Refs.~\cite{Jetzer:1991jr,Schunck:2003kk,Liebling:2012fv,Macedo:2013jja} for reviews), which is a solution to the 
Einstein-Klein-Gordon theory (see action~\eqref{Sekg} below).
Both BSs and oscillatons can arise naturally as the end-state of gravitational 
collapse of scalar fields~\cite{Seidel:1991zh,Garfinkle:2003jf,Okawa:2013jba} and share similar features.
If BSs can form in the universe, they might also form binary systems which would 
be a novel gravitational-wave~(GW) 
source~\cite{Macedo:2013qea,Macedo:2013jja,Giudice:2016zpa,Palenzuela:2017kcg,Cardoso:2016rao,Cardoso:2017cfl}. The 
hypothetical detection of a BS could provide indirect evidence for physics beyond the Standard 
Model~\cite{Barack:2018yly,Sathyaprakash:2019yqt}.

The properties of BSs are strongly related to the scalar 
self-interactions of the model~\cite{Kaup:1968zz,Ruffini:1969qy,Khlopov:1985jw,Seidel:1991zh,
Guth:2014hsa,Brito:2015pxa,Minamitsuji:2018kof,Choi:2019mva}. For a given scalar potential, static BSs form a 
one-parameter 
family of solutions governed by the value of the bosonic field at the center of the star. Similarly to the case of 
perfect-fluid stars --~whose properties depend on the value of the central density~-- the mass displays local 
maxima and minima corresponding to the threshold between branches of stability/instability against radial 
perturbations~\cite{Shapiro:1983du,Lee:1988av,Gleiser:1988rq}.
The maximum mass and compactness of a BS depend strongly on the boson self-interactions. As a rule 
of thumb, the stronger the self-interaction the higher the maximum compactness and mass 
of a stable BS~\cite{Schunck:2003kk,Liebling:2012fv,Macedo:2013jja,Choi:2019mva}. 

The scope of this work is to study a novel class of BSs obtained in the case of a periodic potential [cf. 
Eq.~\eqref{VaxMIO}] inspired by that of the QCD axion~\cite{Peccei:1977hh,Peccei:1996ax}. The standard QCD axion model 
depends on a single parameter (either the axion mass or the axion decay constant, since these two quantities are 
inversely 
proportional to each other). We shall consider an extended version of the parameter space in which the scalar-field mass 
is independent from the decay constant, as in several ALP models~\cite{Irastorza:2018dyq}.
Gravitationally bound configurations of QCD axions (dubbed \emph{axion stars}) have 
been studied in the past few years
\cite{Visinelli:2017ooc,Braaten:2015eeu,Eby:2017xrr,Eby:2016cnq,Helfer:2016ljl,Schiappacasse:2017ham,Chavanis:2017loo,
Clough:2018exo}, mostly in the nonrelativistic limit or considering relativistic collisions in a limited number of 
configurations. One of our goals is to extend these studies to provide a detailed general 
relativistic description of equilibrium solutions and to include the case of an ALP potential in which the axion mass 
and the decay constant are independent from each other.
Since we focus on \emph{complex} scalar fields that oscillate in time, we shall dub our solutions as \emph{axion boson 
stars}~(ABSs). We refer to the Sec.~\ref{sec:discussion} for a discussion about the close connection between complex-
and real-field solutions.

\section{Setup}
\label{Sec1}

\subsection{Axion boson stars in Einstein-Klein-Gordon theory}
\label{EKG}
BSs are equilibrium solutions to Einstein-Klein-Gordon theory\footnote{Henceforth we use $G=c=1$ units and the 
$(-,+,+,+)$ signature.} 
\begin{equation}
S=\int \dd^4 x 
\sqrt{-g}\left[\frac{R}{2\kappa}-\tensor{g}{^{\alpha\beta}}\p\indices{_\alpha}\psi\indices{^*}\p\indices{_\beta}
\psi-V(|\psi|^2)\right]\ ,
\label{Sekg}
\end{equation}
where $\kappa=8\pi$, $R$ is the Ricci curvature, $\psi(x^\mu)$ is a (generically complex) scalar field, and 
$V(|\psi|^2)$ is the scalar self-interaction potential. The field equations are
\begin{eqnarray}
 \tensor{R}{_{\mu\nu}}-\frac{1}{2}\tensor{g}{_{\mu\nu}} R 
 &=&\kappa \left(\p\indices{_\mu} 
\psi\indices{^*}\p\indices{_\nu}\psi+\p\indices{_\nu}\psi\indices{^*}\p\indices{_\mu}\psi-\tensor{g}{_{\mu\nu}}\l[
\tensor{g}{^{\alpha\beta}}\p\indices{_\alpha}\psi\indices{^*}\p\indices{_\beta}\psi+V\r]
\right) \,, 
\label{Einsphi}\\
\dfrac{1}{\sqrt{-g}}\p\indices{_\alpha}\l(\sqrt{-g}\,\tensor{g}{^{\alpha\beta}}\p\indices{_\beta}\psi\r)  
&=&\dfrac{\p\,V}{\p|\psi|^2}\,\psi \,,\label{KGphi}
\end{eqnarray}
together with the complex conjugate of Eq.~\eqref{KGphi}.

We consider spherically symmetric, equilibrium configurations, described by the line element
\begin{equation}
\dd s^2=-e^{v(r)}\dd t^2+e^{u(r)}\dd r^2+r^2\l[\dd \vartheta^2+\sin^2(\vartheta)\dd \varphi^2\r]\ ,
\label{metric}
\end{equation}
in terms of two real metric functions, $v(r)$ and $u(r)$. The scalar field is instead oscillating in time,
\begin{equation}
\psi=\phi(r)\,e^{i\w t}\ .
\label{ansatz}
\end{equation}
where $\omega$ is a real parameter to be determined. The ansatz for the scalar field is required such that the equations of motion are real. With the above ansatz the field equations reduce to
\begin{eqnarray}
 \frac{1}{r^2}\,\frac{\p}{\p r}\l(r\,e^{-u} \r)-\frac{1}{r^2}&=&-\kappa\,\rho\ ,
\label{BS1} \\
e^{-u}\l[\frac{v'}{r}+\frac{1}{r^2}\r]-\frac{1}{r^2}&=&\kappa p_{\rm rad}\ ,
\label{BS2} \\
\phi''+\l[\frac{2}{r}+\frac{v'-u'}{2}\r]\phi'&=&e^{u}\l[U(\phi)-\w^2 e^{-v} \r]\ ,
\label{BS3}
\end{eqnarray}
where $U(\phi)=\dd V(\phi^2)/\dd (\phi^2)$ and we have introduced the energy density $\rho$ and the radial pressure 
$p_{\rm rad}$ which, together with the tangential
pressure $p_{\rm tan}$, are expressed in terms of the stress-energy tensor of the scalar field, $ T_{\alpha\beta}$, 
as follows
\begin{eqnarray}
 \rho&\equiv&-\tensor{T}{^0_0}=\w^2e^{-v}\phi^2+e^{-u}{\phi'}^2+V(\phi)\ , \label{rho}\\
 p_{\rm rad}&\equiv& \tensor{T}{^1_1}=\w^2e^{-v}\phi^2+e^{-u}{\phi'}^2-V(\phi)\ , \label{prad}\\
 p_{\rm tan}&\equiv& \tensor{T}{^2_2}=\w^2e^{-v}\phi^2-e^{-u}{\phi'}^2-V(\phi)\ . \label{ptan}
\end{eqnarray}
Unlike the case of perfect fluid stars, the scalar field introduces anisotropies, $p_{\rm rad}\neq p_{\rm tan}$.

In the following we shall focus on a specific form of the 
potential, motivated by that of the QCD axion. The latter acquires a periodic potential due to non-linear 
instanton effects (for a detailed derivation, see Ref.~\cite{diCortona:2015ldu}). We shall 
consider the following extension of the QCD axion potential:
\begin{equation}
V(\phi)=\frac{2\mu^2_{a} \,f_{a}^2}{\hbar B}\l(1-\sqrt{1-4\,B\,\sin^2\l(\dfrac{\phi\sqrt{\hbar}} { 
2\, f_{a} } \r)}\r)\ ,
\label{VaxMIO}
\end{equation}
where $B=\frac{z}{(1+z)^2}\approx0.22$ and $z\equiv m_{u}/m_{d}\approx0.48$ is the mass
ratio of the up and down quarks. 
The second term in the above potential is the standard QCD axion potential~\cite{diCortona:2015ldu}, to which we added 
the first constant\footnote{A constant term is innocuous is flat spacetime but leads to asymptotically (anti) de Sitter 
solutions when the scalar field is coupled to gravity.} term to ensure $V(0)=0$ and hence asymptotically flat solutions.

In the above potential, $\mu_a$ and $f_a$ are two free parameters, with dimensions of an 
inverse length and of an energy, respectively. By expanding Eq.~\eqref{VaxMIO} 
around the minimum at $\phi=0$, we obtain
\begin{equation}
 V(\phi)\sim \mu_a^2 \phi^2-\left(\frac{3B-1}{12}\right)\frac{\hbar\mu_a^2}{f_a^2}\phi^4+{\cal 
O}(\phi^6)\,, 
\label{Vexp}
\end{equation}
from which we can identify the ALP mass 
\begin{equation}
 m_a = \mu_a \hbar\,,
\end{equation}
and the ALP quartic coupling. In this model $\mu_a$ and $f_a$ are independent parameters. The case of a QCD axion is 
included when~\cite{diCortona:2015ldu} 
\begin{equation}
 m_a\approx 5.7\,\mu{\rm eV}\left(\frac{10^{12}\,{\rm GeV}}{f_a}\right)\,. \label{axionmass}
\end{equation}

Note that the expansion in Eq.~\eqref{Vexp} is valid when $f_a\gg \phi\sqrt{\hbar}$. In this limit, to the leading 
order, the potential reduces to the standard mass term, which leads to the family of so-called ``mini'' 
BSs~\cite{Kaup:1968zz,Ruffini:1969qy}.

\subsection{Numerical method}
\label{NumMeth}
The field equations~\eqref{BS1}-\eqref{BS3} can be solved for 
$v(r)$, $u(r)$ and $\phi(r)$ by imposing regularity boundary conditions at the center
\begin{equation}
 \phi(0)=\phi_c\,,\quad u(0)=0\,,\quad v(0)=v_c\,,
\end{equation}
and asymptotic flatness
\begin{equation}
\lim_{r\to\infty}\phi=0\,,\quad \lim_{r\to\infty}v=0\,.
\end{equation}
The value of $v_c$ is arbitrary since it can be set to unity by a rescaling of the time coordinate.

For any choice of the central value of the scalar, $\phi_c$, the above conditions define a boundary value problem, 
whose eigenvalues $\omega=\omega_m$ form an infinite discrete set $(m=0,1,2,...$). We solve numerically the 
one-dimensional boundary-value 
problem through a shooting method~\cite{Shooting,Press:1992zz}, following the same procedure detailed in 
Ref.~\cite{Macedo:2013jja}. For a given $\phi_c$, different eigenvalues correspond to the fundamental mode 
--~for which the corresponding scalar eigenfunction has zero radial nodes ($m=0$)~-- and to excited states --~with one 
($m=1$) or more ($m>1$) nodes. 
Excited states are expected to be unstable and therefore we focus only on the fundamental mode~\cite{Lee:1988av}.

As in the case of ordinary fluid stars, we can also characterize our solutions according to their 
total mass $M$ and radius $R$. Both quantities can be defined through the mass function ${\cal M}(r)$, given by
\begin{equation}
{\cal M}(r)=\frac{r}{2}\l[1-e^{-u(r)}\r]\ .
\label{MASSA}
\end{equation}
Then, the total mass is $M={\cal M}(r\to\infty)$. At distances much larger than the Compton wavelength of the scalar 
field, the latter is exponentially suppressed but it has nonetheless support up to infinity. Therefore, there is no 
real vacuum for BS spacetimes. For this reason, the radius of a BS is not uniquely defined.\footnote{An exception are 
BSs with V-shaped potentials~\cite{Hartmann:2012da}.} It is customary to define an effective radius, $R$, as that 
containing $99\%$ of the mass, i.e., ${\cal M}(R)=0.99M$.
Note that very compact BSs might have a very steep scalar profile, practically removing the ambiguity in the definition 
of the radius~\cite{Macedo:2013jja}.

\subsection{Stability analysis}
\label{stab}

\begin{figure*}
  \includegraphics[width=0.46\textwidth,angle=0,keepaspectratio]{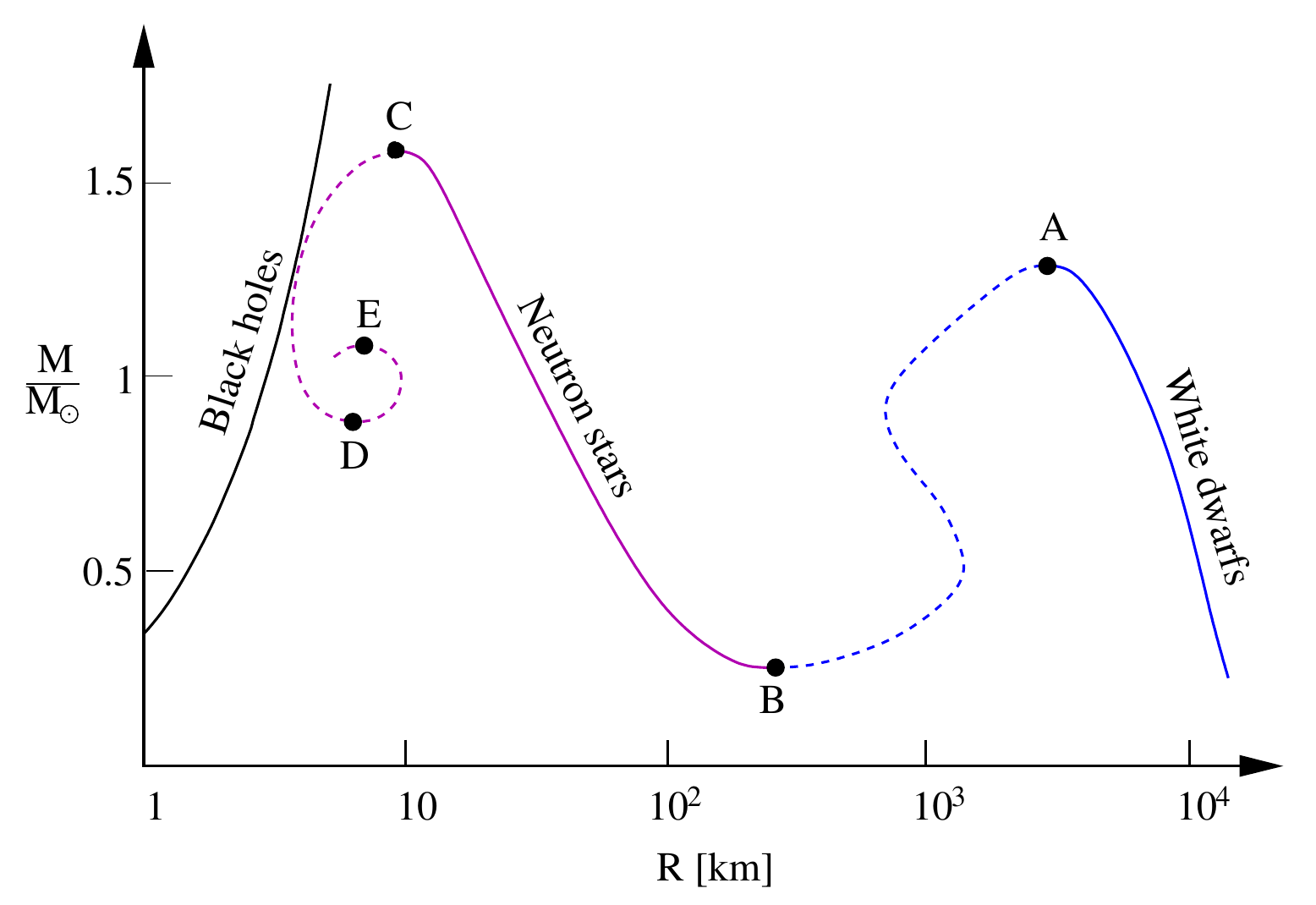}
  \includegraphics[width=0.52\textwidth,angle=0,keepaspectratio]{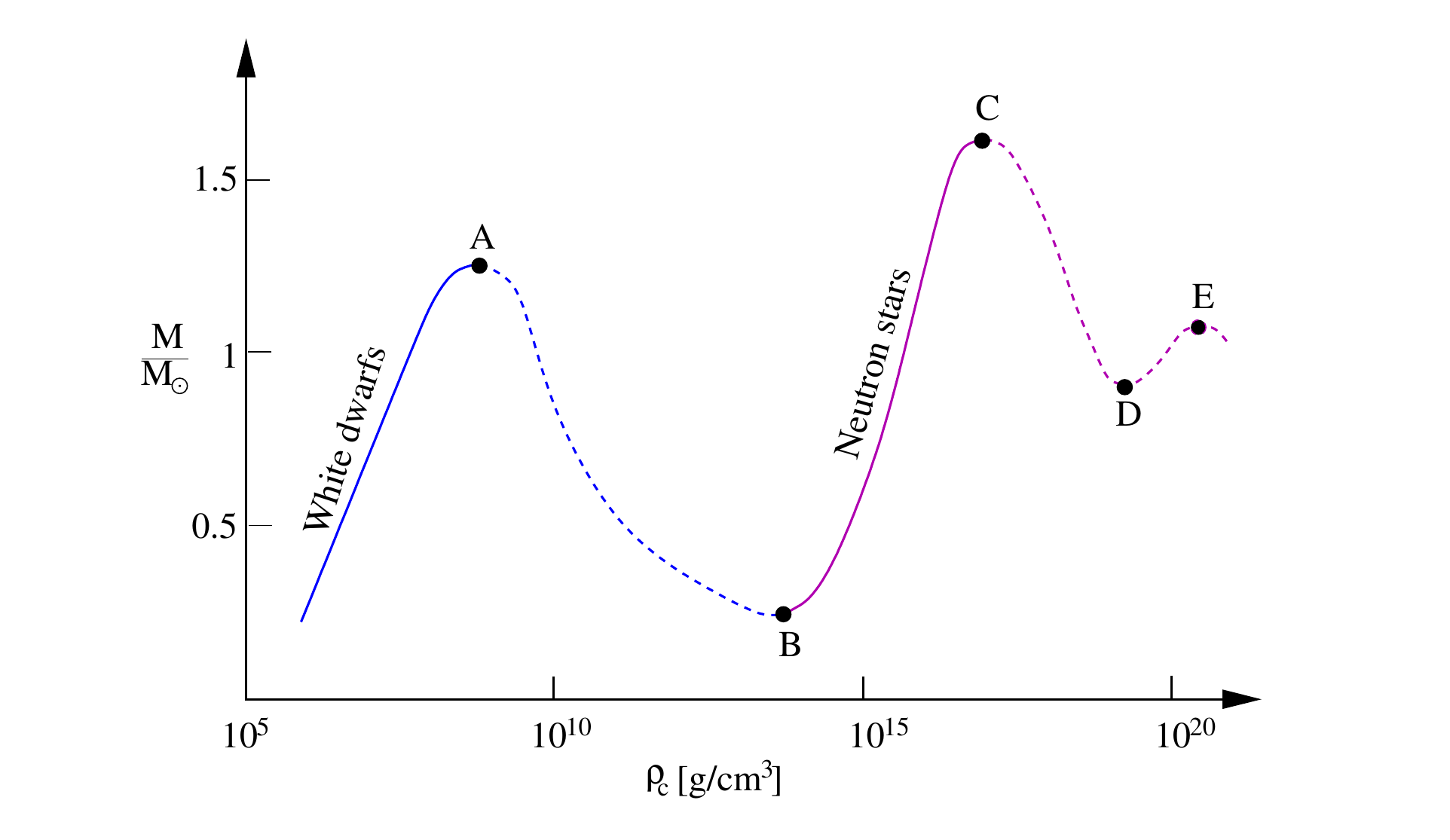}
\caption{Schematic diagrams of the mass-radius (left) and mass-central 
density (right) relations for ordinary compact stars. The 
continuous (dotted) curves represent stable (unstable) equilibrium configurations. In the left diagram, the black curve 
represents the black-hole case, $R=2M$. Taken from Ref.~\cite{Kleihaus:2011sx} (adapted from \cite{Shapiro:1983du}).}
\label{fig:stability}
\end{figure*}

The radial stability analysis of a BS can be performed by looking at the $M(\phi_c)$ and $M(R)$ diagrams, similarly 
to the case of perfect-fluid stars~\cite{Shapiro:1983du,Lee:1988av,Gleiser:1988rq,ChavanisB}. 
A schematic representation of these diagrams for perfect-fluid stars is shown in 
Fig.~\ref{fig:stability}.
The critical points are those for which $dM/d\rho_c=0$ and correspond to a change of sign of a given radial 
mode, $w_n$. The stability can be studied by recalling that the modes are ordered, i.e. $w_0^2<w_1^2<w_2^2<...$, and 
that $dR/d\rho_c>0$ (resp. $dR/d\rho_c<0$) corresponds to 
the change of sign of a mode with odd (resp. even) value of $n$~\cite{Shapiro:1983du}. 

Thus, if the left-most branch of the $M(\rho_c)$ diagram is stable (``white dwarf'' branch, $w_n^2>0$ for any $n$), the 
first point at which $dM/d\rho_c=0$ (Point A in Fig.~\ref{fig:stability}) corresponds to $dR/d\rho_c<0$ and therefore 
$w_0^2$ 
changes sign, leading to an instability in the AB branch. In Point B $dR/d\rho_c<0$ so $w_1^2$ cannot change sign, nor 
can $w_2^2$ become negative since it must be larger than $w_1$. Therefore the 
only option is that $w_0^2$ returns positive, bringing back stability in the BC branch (``neutron star'' branch). 
Following the same argument, it is straightforward to see that all configurations after Point~C are unstable. This is 
due to the fact that $dR/d\rho_c$ is alternatively positive and negative at the extrema points~D,~E and so on, 
corresponding to the typical ``curl'' of the $M(R)$ diagram shown in Fig.~\ref{fig:stability}. 
Therefore, standard perfect-fluid stars beyond the stable neutron-star branch are unstable 
and bound to either collapse to a black hole or to migrate to a stable stellar configuration by losing mass.

As we shall discuss, the same analysis can be performed for our ABS configurations, with the central scalar 
field $\phi_c$ playing the same role of the central mass density $\rho_c$~\cite{Lee:1988av,Gleiser:1988rq}. In contrast 
with many BS models, ABSs can present many stable and unstable branches depending on the parameter~$f_a$.

\section{Results}
Here we present the properties of spherically-symmetric ABSs by 
solving Eqs.~\eqref{BS1}-\eqref{BS3} with the aforementioned boundary conditions and assuming the potential given in 
Eq.~\eqref{VaxMIO}.

For the purpose of the numerical integration, it is convenient to use dimensionless quantities by performing the 
following rescaling
\begin{equation}
r\to\frac{\tilde{r}}{\mu_{a}}\ ;\quad {\cal M}(r)\to\frac{\tilde{{\cal M}}(\tilde{r})}{\mu_{a}}\ ;\quad \w\to\tilde{\w}\,\mu_{a}\ 
;\quad \phi(r)\to\frac{\tilde{\phi}(\tilde{r})}{\sqrt{4\pi}}\ ;\quad f_{a}\to\tilde{f}_{a} 
\sqrt{\hbar}\ , \label{rescaling}
\end{equation}
where all tilded variables are dimensionless. (Henceforth we shall also rescale the radius and the mass as $\tilde 
R= R\mu_s$ and $\tilde M=R \mu_s$.)
The rescaled field equations that we integrated numerically are given in Appendix~\ref{app:eqs} for completeness. 
Notice that the parameter $\mu_a$ disappears from the equations after the rescaling, so effectively any dimensionful 
quantity is written in units of $\mu_a$. This also implies that the mass and the radius of a given configuration scale 
as $1/\mu_a$ so they can assume any value depending on $\mu_a$. The compactness of the star, $M/R$, is instead 
independent of $\mu_a$ and therefore fixed for a given configuration.

\begin{table}[t]
\centering
\caption{Representative examples of ABS configurations for $\tilde f_a=10$ (i.e., $f_{a}\approx1.22\times 
10^{20}\,\text{GeV}$). Configurations I and II correspond to that of maximum mass and that of maximum compactness. 
Since in this case $\tilde f_a\gg1$, the results are in very good agreement with the mini BS 
case~\cite{Macedo:2013qea}.}
\label{tabax}
\vspace{0.25cm}
\begin{tabular}{c|c|c|c|c|c}
\hline
\hline
Configuration & $\tilde{\phi}_c$ & $\tilde{\omega}$ & $\tilde{M}$ & $\tilde{R}$ & 
$\tilde{R}/\tilde{M}$ 
\\ \hline 
I     & $0.19150$        & $0.85314$        & $0.63299$              & $7.86381$               & 
$12.4231$       
      \\ \hline
II     & $0.43550$        & $0.77018$        & $0.51725$              & $4.6588$                & 
$9.00685$      \\
\hline
\hline
     
\end{tabular}
\end{table}

As a check of our numerical method, we first integrate the equations for large values of $\tilde f_a$, which should 
reduce to the mini BS case. The large-$\tilde f_a$ limit is quickly saturated and $\tilde f_a\sim {\cal O}(10)$ is 
already in the asymptotic regime. In Table~\ref{tabax} we show two 
representative configurations for 
\begin{equation}
\tilde{f}_{a}=10\quad \Rightarrow \quad f_{a}\approx 1.22\times 10^{20}\,\text{GeV}\ .
\end{equation}
which are in very good agreement with the mini BS case~\cite{Macedo:2013qea}, as expected.
The above value of the decay constant corresponds to a mass of the QCD axion $m_a\approx 5.6\times 10^{-14}\,{\rm eV}$ 
[cf. Eq.~\eqref{axionmass}], which would give a maximum BS mass $M_{\rm max}\approx 1049\,M_\odot$ [cf. 
Table~\ref{tabax}].

\begin{figure}
	\centering
	\includegraphics[width=1\textwidth]{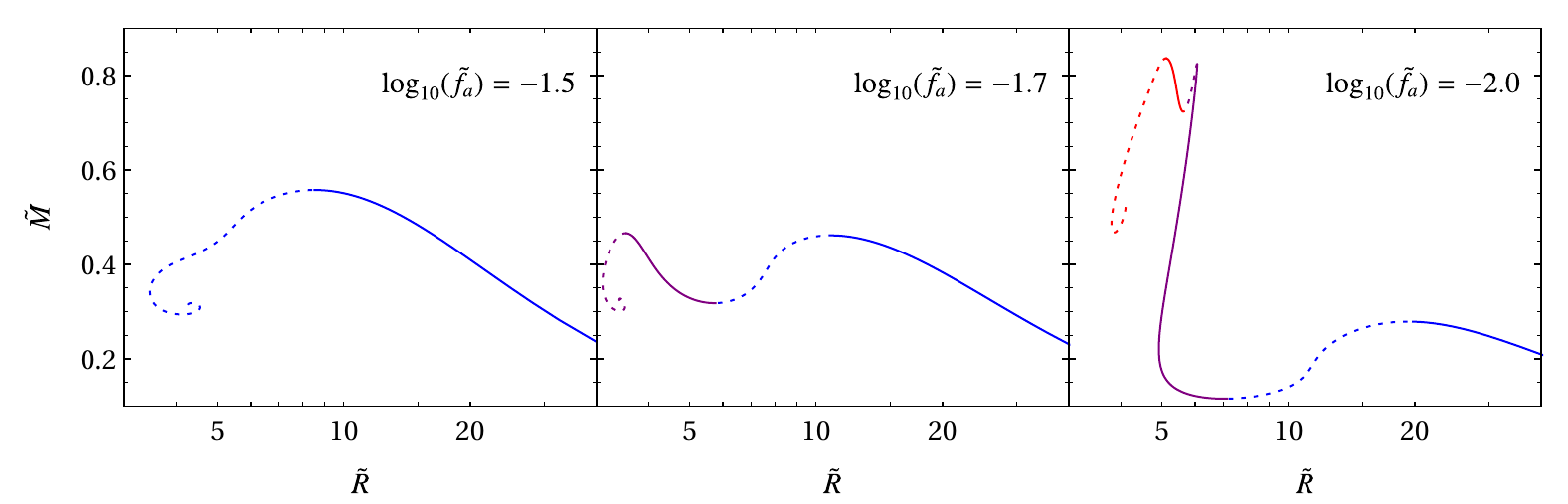}\\
	\includegraphics[width=1\textwidth]{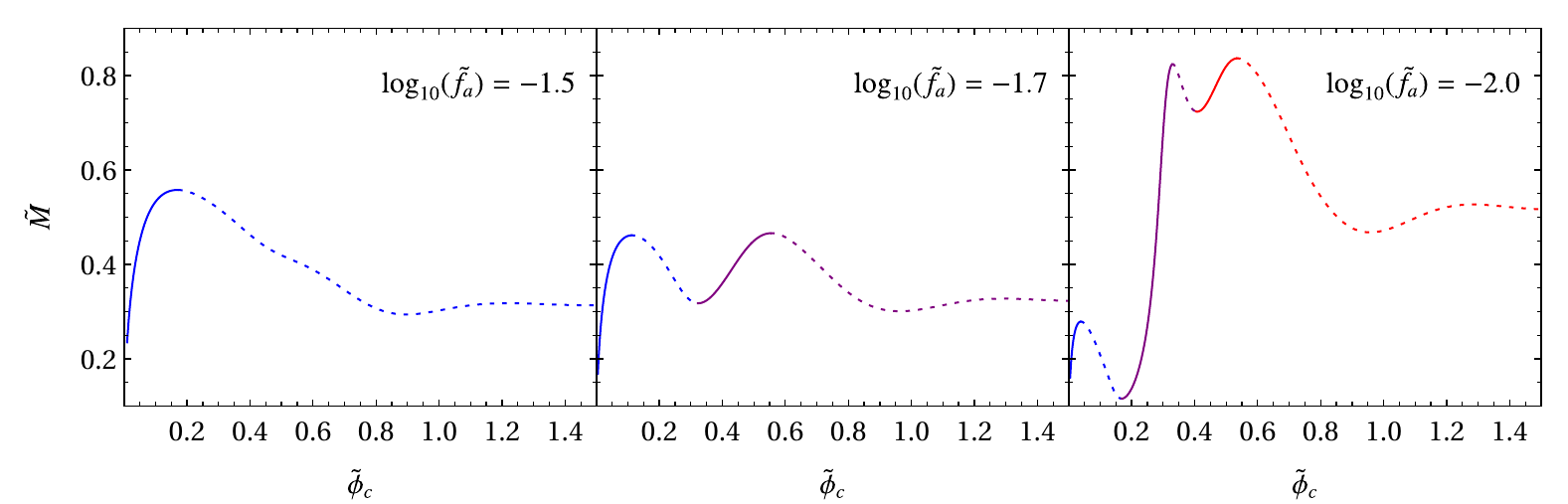}
	\caption{Mass-radius (upper panels) and mass-$\phi_c$ (lower panels) relation for ABSs for 
		different values of $f_a$. Solid (dashed) curves correspond to stable (unstable) configurations. 
		Different colors correspond to different branches. Decreasing the value of the 
		energy scale $\tilde f_a$ causes the appearance of more stable/unstable branches. 
		} 
	\label{final}
\end{figure}

In Fig.~\ref{final} we show the $M-R$ (upper panels) and $M-\tilde\phi_c$ (lower panels) diagrams for three 
representative cases. In the left panels we consider a relatively large value of $\tilde f_a$, which behaves similar to 
the mini BS limit. As expected the model displays a single stable branch up to a critical value of $\tilde\phi_c$, 
beyond which all configurations are unstable.
Remarkably, the decrease of the decay constant leads to the formation of several new branches. 
As $\tilde f_a$ decreases, the $M-\tilde\phi_c$ diagram starts developing multiple maxima, which correspond to various 
turning points in the $M-R$ diagram, as shown Fig.~\ref{final}. A radial stability analysis shows that new 
\emph{stable} 
branches appear, which might correspond to very compact configurations.
In the middle panels of Fig.~\ref{final} we show the case of $f_a\approx 2.4\times 10^{17}\,{\rm GeV}$, which leads to 
the formation of \emph{two} stable branches, similarly to the fluid star case shown in Fig.~\ref{fig:stability}, and 
also to the case of non-relativistic axion stars~\cite{Braaten:2015eeu}, and of solitonic 
BSs~\cite{Macedo:2013jja}.
In the right panels of Fig.~\ref{final} we show the case $f_a\approx 1.2\times 10^{17}\,{\rm GeV}$, which leads to 
the formation of \emph{three} stable branches. This is possible because --~at variance with the previous case and with 
the perfect-fluid case shown in Fig.~\ref{fig:stability}~-- the second minimum of $M(\phi_c)$ occurs when 
$dR/d\rho_c<0$, which in turn depends on the peculiar ``turn'' made by the second branch in the top-right panel of 
Fig.~\ref{final}.

\begin{figure*}
	\includegraphics[width=\linewidth]{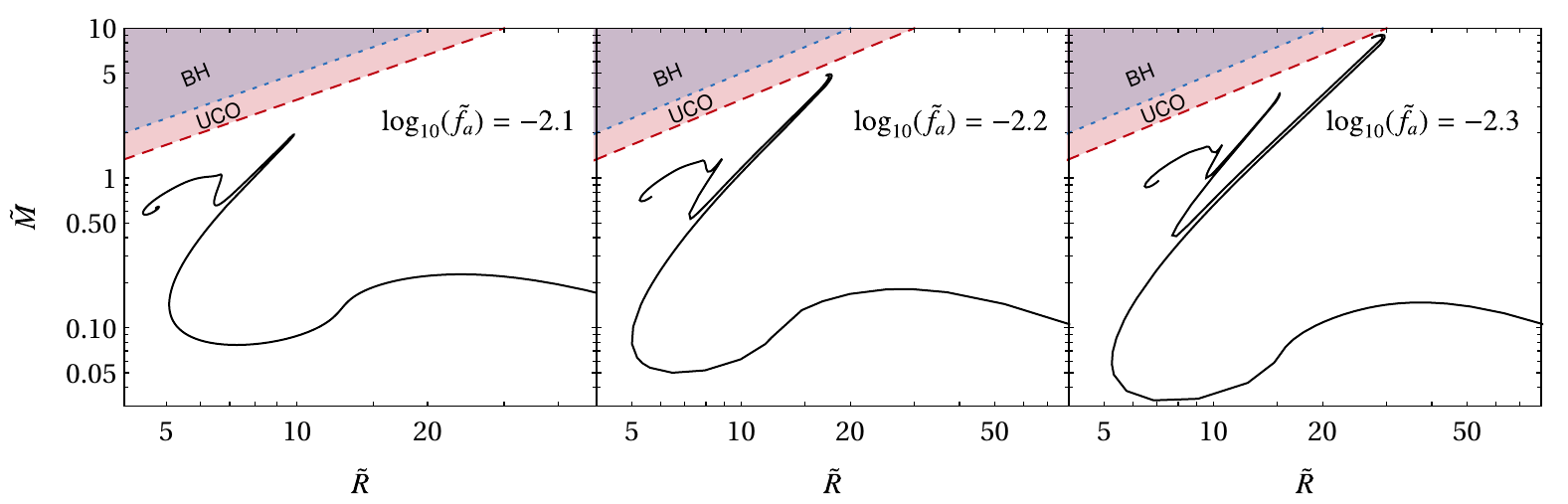}\\
	\includegraphics[width=\linewidth]{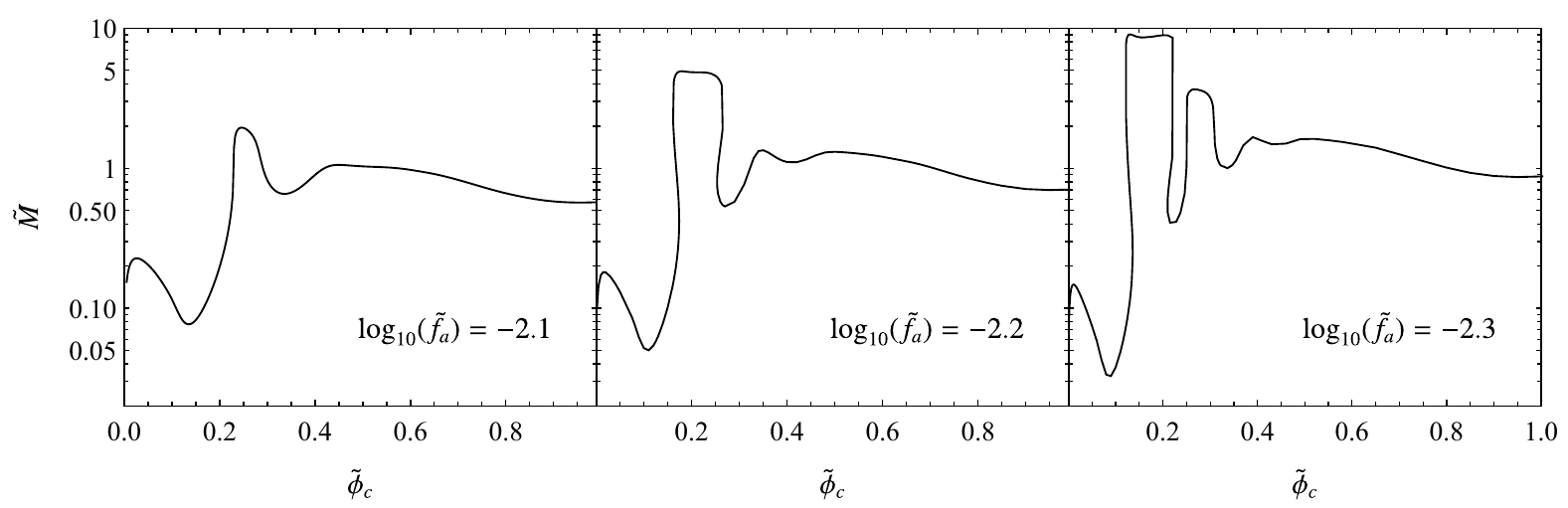}
	\caption{Same as in Fig.~\ref{final} but for a representative sample of ABSs with smaller values of 
$\tilde f_a$. Decreasing the value of $f_a$ leads to the appearance of more 
stable/unstable branches. Additionally, highly compact configurations, with $R\lesssim 3 M$, also appear. The dotted 
and dashed straight line correspond to the black-hole case, $R=2M$, and to the photon-sphere radius, $R=3M$, which 
define the ``ultracompact objects''~(UCOs) introduced in Refs.~\cite{Cardoso:2017cqb,Cardoso:2019rvt}. Even if 
$R>3M$ for all configurations, some of the most compact ones feature a photon-sphere and therefore classify as UCOs.
}
	\label{fig:highcomp}
\end{figure*}

This peculiar behavior keeps occurring as $\tilde f_a$ decreases further. In this case the diagram becomes 
increasingly more intricate and highly sensitive to the value of $\tilde f_a$. This is shown in Fig.~\ref{fig:highcomp} 
for highly compact ABSs. As $\tilde f_a$ decreases, it becomes increasingly more 
difficult to find a solution, as the corresponding eigenvalue requires an extreme fine tuning\footnote{A similar 
numerical issue was encountered for so-called ``solitonic BSs'' in Ref.~\cite{Macedo:2013jja}.}. 
Thus, we were able to construct complete diagrams down at most to $\tilde f_a=10^{-2.3}$, corresponding to $f_a\approx 
6\times10^{16}\,{\rm GeV}$, i.e. close to the 
GUT scale. 

As shown in Fig.~\ref{fig:highcomp}, as $f_a$ decreases the most compact configurations approach the compactness 
$M/R\sim 1/3$. We expect that the compactness of such solutions increases when the value of $f_a$ 
decreases, i.e., 
that the curves in Fig.~\ref{fig:highcomp} approach more the UCO region~\cite{Cardoso:2017cqb,Cardoso:2019rvt} (shaded 
region).  Due to the numerical limitations of our code, we were unable to further explore the mass-radius diagram for 
smaller values of $f_a$. However, we performed a numerical search and we indeed confirm that the compactness decreases 
for smaller $f_a$, although always satisfying the bound $M/R>1/3$.
Thus, the effective radius is always outside the Schwarzschild photon-sphere. However, since the radius of these 
solutions cannot be defined uniquely, we searched for light rings in our numerical solutions by computing the effective 
potential for null particles~\cite{Cardoso:2014sna}. In 
regular, horizonless objects, due to the structure of the effective potential and to the centrifugal barrier at 
the center of the star, light-rings come in pair: an inner stable one and an outer unstable 
one~\cite{Cardoso:2014sna,Cunha:2017qtt}. We found that a pair of light-rings appear for approximately 
$\log_{10}(\tilde f_a)\leq-2.34$. Although it is numerically challenging to find ultracompact solutions with significantly smaller values of 
$f_a$, we expect that light-rings exist generically in this regime.
Therefore, even if $R>3M$ for all configurations found in this work, the most compact ones --~existing for 
$f_a\lesssim 6\times 10^{16}\,{\rm GeV}$~-- classify as UCOs~\cite{Cardoso:2017cqb,Cardoso:2019rvt}.

\begin{figure}[h!]
	\centering
	\includegraphics[width=0.48\textwidth]{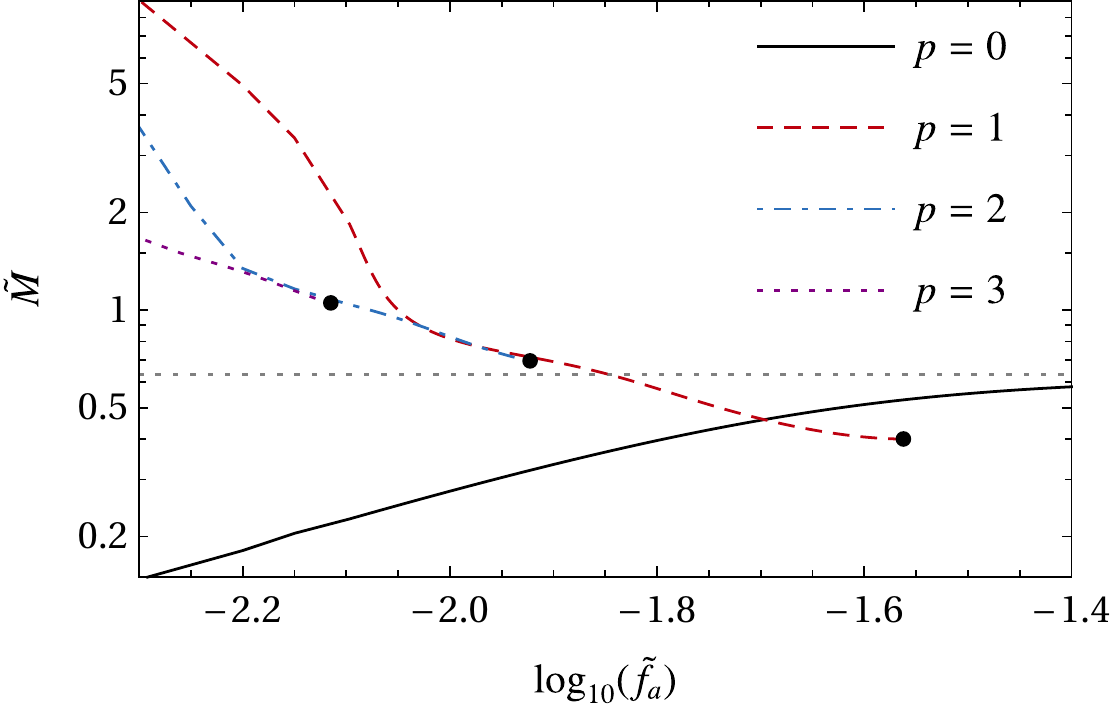}
	\includegraphics[width=0.48\textwidth]{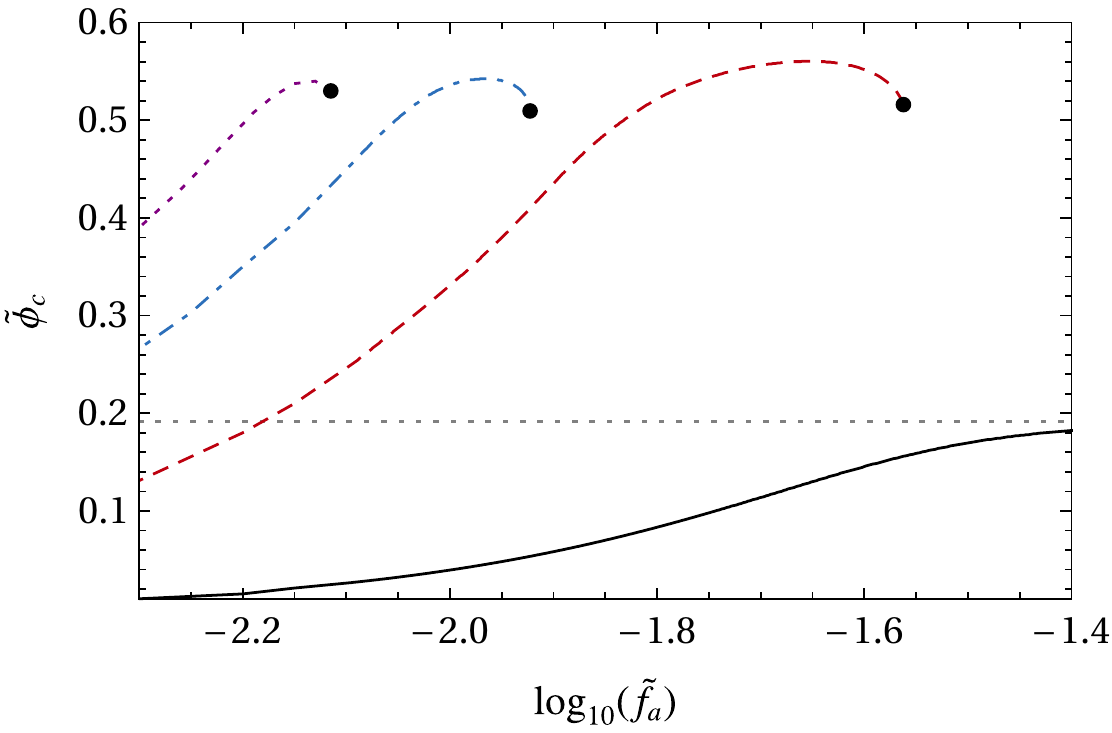}
	\includegraphics[width=0.48\textwidth]{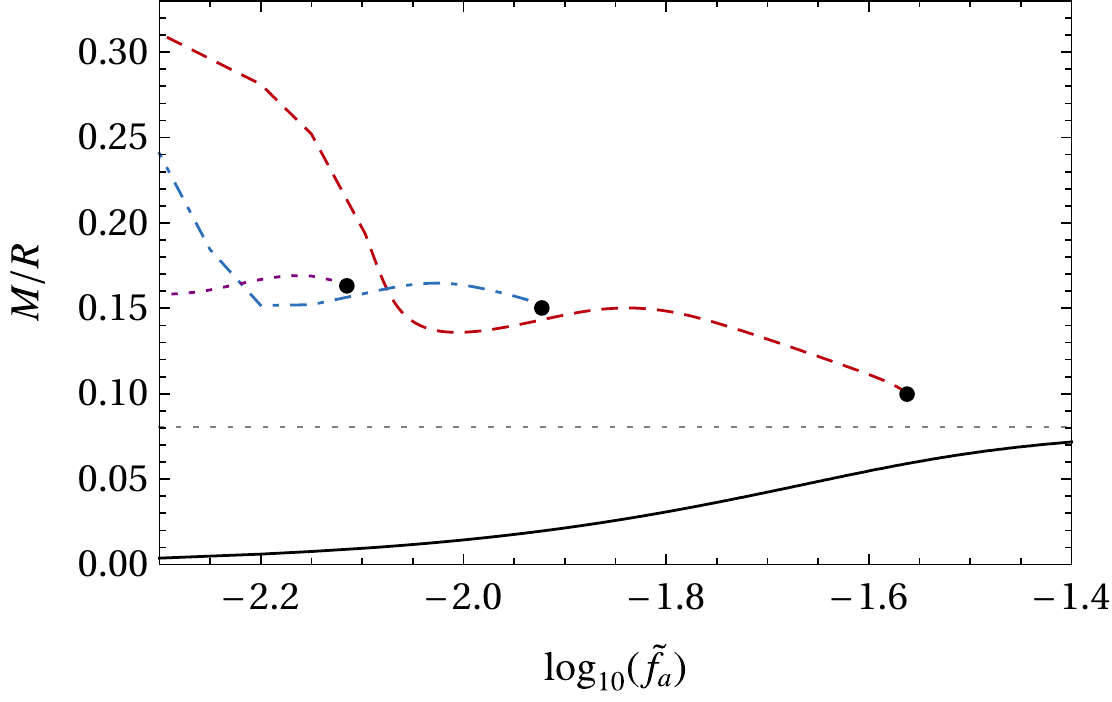}
	\caption{Mass (top-left), scalar field (top-right) and compactness (bottom) of the maximum mass MSCs
as function of the axion decay constant. The dotted horizontal lines represent the mini BS limit and the markers 
indicate the appearance of additional branches at $\tilde f_a=\tilde f_{a}^{\rm crit}$. As $f_a$ decreases, more stable 
branches appear. The maximum mass of the first branch --~the one with the BS limit (dotted line)~-- decreases as 
$f_a$ decreases, while the mass of the other branches increases. In the plots we track the first three additional branches, but subsequent branches appear for lower 
values of $\tilde f_a$ (three additional branches are shown in Table~\ref{tab:bran})
}
	\label{fig:branch}
\end{figure}

We notice that the appearance of additional branches continues for lower values of $f_a$. Each new branch corresponds 
to the appearance of two additional marginally stable configurations~(MSCs) occurring at the maximum and minimum 
of the diagrams (see Fig.~\ref{final}). In 
Fig.~\ref{fig:branch} we track the MSCs with maximum masses as function of $f_a$. We notice that 
while the mass of the first MSC --~the one that has mini BS as limiting 
case~-- decreases as $f_a\to 0$, there are other MSCs 
whose masses increase quickly in the same limit. These solutions are responsible for the appearance of stable UCOs, 
discussed previously. Another interesting feature of these secondary MSCs is that they seem to appear around the same 
value of the central scalar field
$\tilde\phi_c\sim 0.52$, and for critical values of the decay constant corresponding to the minima of the potential 
(where the latter also vanishes), i.e.
\begin{equation}
\tilde{f}_{a}^{\rm crit}\sim \frac{0.52}{4 p \pi^{3/2}}\,,
\label{eq:branch}
\end{equation}
with $p\geq 1$ being an integer label for the additional branches. In Table \ref{tab:bran}, we compare the
numerical computation of $\tilde f_{a}^{\rm crit}$ with Eq.~\eqref{eq:branch} for the first six additional branches. We see that the agreement improves for 
lower values of $f_a$. It is not clear up to which point this behavior (or the appearance of new branches) will hold 
for even lower values of $f_a$, but we expect this general trend of the solutions.

\begin{table}
\centering
\caption{Comparison between the numerically computed $\log_{10}(\tilde f_{a}^{\rm crit})$ and the analytical expression 
given by Eq.~\eqref{eq:branch}.}
\begin{tabular}{c|c|c}
	$p$ &  Numerical & Eq.~\eqref{eq:branch}\\
\hline
	1 & -1.562 & -1.6318 \\\hline
	2 & -1.922 & -1.9328 \\\hline
	3 & -2.115 & -2.1089\\\hline
	4 & -2.242 & -2.2338 \\\hline
	5 & -2.338 & -2.3307 \\\hline
	6 & -2.416 & -2.4099 \\
\end{tabular}
\label{tab:bran}
\end{table}


\section{Discussion}
\label{sec:discussion}

We were able to construct solutions with a decay constant as small as $f_a\approx 6\times10^{16}\,{\rm GeV}$, i.e. 
close to the GUT scale. This corresponds to a QCD axion mass $m_a\approx 9.3\times 10^{-11}\,{\rm eV}$. 
It is worth noting that this value of the QCD axion mass evades the constraints coming from the black-hole superradiant 
instability~\cite{Arvanitaki:2010sy,Arvanitaki:2014wva,Brito:2015oca,Brito:2017wnc,Brito:2017zvb,Cardoso:2018tly} 
for two reasons: first because the mass is slightly larger than what probed by superradiance (smaller values of $f_a$ 
would correspond to even larger masses, for which the superradiant constraints do not apply), and second because the 
corresponding value of the decay constant would give a relatively strong axion-photon coupling, potentially 
quenching the superradiant instability~\cite{Ikeda:2019fvj,Boskovic:2018lkj}. 

In this unconstrained region of the parameter space~\cite{Irastorza:2018dyq}, GW observations of isolated 
or binary ABSs might provide indirect evidence for ultralight bosonic fields in the universe.
From the right top panel of Fig.~\ref{fig:highcomp}, an axion with mass $m_a\approx 9.3\times 
10^{-11}\,{\rm eV}$ and $f_a\approx6\times10^{16}\,{\rm GeV}$ would correspond to an ABS with maximum mass $M_{\rm 
max}\approx 9\, M_\odot$, and with a radius only slightly larger that of the photon-sphere, i.e. $M/R\lesssim1/3$.
The signal from the coalescence of such two ABSs would be very similar to that from two black holes with the 
same mass, but with some distinctive features~\cite{Cardoso:2019rvt}: (i) if the ABSs are spinning, their 
quadrupole moment would differ from that of a Kerr black hole~\cite{Krishnendu:2017shb}; (ii) even when the binary 
is nonspinning, tidal heating is expected to be negligible for BS binaries while it might affect a black-hole binary in 
a detectable way~\cite{Maselli:2017cmm}; (iii) at variance with black holes, BSs can be tidally deformed in the last 
stages of the inspiral, similarly to neutron stars~\cite{Cardoso:2017cfl,Sennett:2017etc}. These three effects would 
leave an imprint in 
the inspiral (at various post-Newtonian orders) compared to the case of a black-hole binary. Furthermore, the merger of 
two ABSs might be much richer than that of two black holes. For instance, scalar modes might be excited in the 
post-merger phase, and the merger remnant might be either a distorted ABS or a black hole, depending on the mass of the 
binary components~\cite{Liebling:2012fv,Palenzuela:2017kcg}. 
The mass-radius diagram in the right-top panel of Fig.~\ref{fig:highcomp} suggests that, when the 
decay constant is at the GUT scale, the merger of two ABSs with $M\approx 4\,M_\odot$ (corresponding to a 
compactness $M/R\approx 0.2$) can result in a \emph{radially stable} ABS. The corresponding GW signal falls in the band 
of ground-based detectors.
Furthermore, even for binaries with total mass exceeding the maximum mass of a given model, the presence of several 
stable branches makes it more difficult to predict the final state. Indeed, we expect multiple outcomes for the merger 
remnant, depending on the initial conditions.
We therefore expect that inspiral-merger-ringdown consistency tests of the entire coalescence 
signal~\cite{TheLIGOScientific:2016src} 
can provide a strong discriminator between ABSs and black-hole binaries, similarly to case of two 
highly-compact ``solitonic'' BSs~\cite{Palenzuela:2017kcg}.

Furthermore, for an ALP field with $m_a\lesssim 10^{-16}\,{\rm eV}$, the corresponding ABS is supermassive, $M\gtrsim 
10^{6}\, M_\odot$. While this is true for all BS models, in the ABS case when $f_a$ is sufficiently small the 
corresponding star is very compact and can feature a photon sphere. It would be therefore interesting to constrain this 
model with the recent shadow observation of the Event Horizon Telescope~\cite{Akiyama:2019cqa,Olivares:2018abq}. In this 
case, isolated and binary supermassive ABSs would also be an exotic target of the LISA mission~\cite{Audley:2017drz} 
and of evolved concepts thereof~\cite{Baibhav:2019rsa}. 

Although the most compact ABS configurations possess two light rings, our numerical results lead us to conjecture 
that the compactness of stable ABSs approaches the photon-sphere limit, $M/R=1/3$, but 
never exceeds it. This is possible because the geometry near the effective surface of the ABS is not Schwarzschild, and 
therefore a light ring might appear even when $R\gtrsim 3M$. It would be interesting to test our conjecture by 
decreasing the value of $\tilde f_a$ further and, possibly, to find an analytical explanation for the maximum value of 
the compactness of ABSs.
Probing smaller values of $f_a$ is a natural technical extension of our work, which might be better achieved using 
relaxation methods~\cite{Press:1992zz}, since the latter might be more efficient to find highly-compact solutions than 
the shooting method.

Another relevant follow-up of our work concerns the non-radial stability of ABSs, which requires a proper 
linear-perturbation analysis. In case such solutions turn out to be linearly stable, it would be interesting to study 
whether they can form in ABS collisions~\cite{Helfer:2016ljl,Palenzuela:2017kcg,Clough:2018exo}, or what would be the 
outcome of the coalescence of two highly-compact ABSs~\cite{Helfer:2016ljl,Palenzuela:2017kcg}.

For simplicity we focused on the case in which the scalar field is complex and oscillating in time. This gives a 
\emph{static} stress-energy tensor and therefore a static metric, at variance with the case of 
oscillatons~\cite{Seidel:1991zh}. Both BSs and oscillatons can be formed as the 
end-state of gravitational collapse of scalar fields~\cite{Seidel:1991zh,Garfinkle:2003jf,Okawa:2013jba}, and both 
configurations share similar features. Therefore, we expect that the properties discussed above for ABSs would be 
qualitatively the same also for axion stars made of a real field, which is the most relevant case for ALPs and the QCD 
axion. 

Alternatively, due to the similarities of the solutions presented here with the solitonic BSs~\cite{Lee:1991ax}, it 
would be interesting to compare the general relativistic case with calculations in absence of gravity (i.e., considering 
Q-balls~\cite{Coleman:1985ki} or oscillons~\cite{Bogolyubsky:1976nx}). These solutions have been receiving considerable 
attention, and cases with oscillatory potentials similar to the one adopted here were explored 
recently~\cite{Mukaida:2016hwd,Eby:2018ufi}. The existence of a Q-balls and oscillons with the same self-interacting 
potential could be an indication that these solutions might describe ultracompact objects when gravity is present.

Finally, BSs made of complex massive vector fields (so-called Proca stars) also exist and have properties similar to 
their scalar counterparts~\cite{Brito:2015pxa,Minamitsuji:2018kof}. We expect that Proca stars with a self-interaction 
potential similar to the one assumed in this work would share the same qualitative features of ABSs.


\section*{Acknowledgments}
We acknowledge support provided under the European Union's H2020 ERC, Starting 
Grant agreement no.~DarkGRA--757480 and by the Amaldi Research Center funded by the MIUR program "Dipartimento di 
Eccellenza" (CUP: B81I18001170001).
C.F.B.M. thanks the Conselho Nacional de Desenvolvimento Cient\'ifico e Tecnol\'ogico (CNPq) for partial financial support.
\vskip 2cm

\appendix

\section{Field equations in dimensionless form}\label{app:eqs}

For completeness we give here the field equations that we integrated numerically --after the 
rescaling~\eqref{rescaling}. Setting $\tilde r = x$ and $W(x)=e^{v(x)}$, the final equations are

\be
\begin{split}
\tilde{{\cal M}}'(x)-\frac{x}{B\,W(x)}\l[B\,W(x)\l(x-2\tilde{{\cal M}}(x)\r)\tilde{\phi}'(x)^2+B\,x\,\w^2\tilde{\phi}
_{a}^2(x)\r.\\
\l.+8\pi\,x\,\tilde{f}_{a}^2\,W(x)\l(1-\sqrt{1-2\,B+2\,B\,\cos\l(\frac{\tilde{\phi}}{2\sqrt{\pi}\,\tilde{f
}_{a}}\r)}\r)\r]=0\ ,\\
\end{split}
\ee
\be
\begin{split}
B\,x\l(x-2\,\tilde{{\cal M}}(x)\r)W'(x)-\l[\l(2\,B\,x^3-4\,B\,x^2\,\tilde{{\cal M}}(x)\r)\tilde{\phi}'(x)^2 \r.\\
\l.-16\pi\,f_{a}^2\,x^3\l(1-\sqrt{1-2\,B+2\,B\,\cos\l(\frac{\phi(x)}{2\sqrt{\pi}\,f_{a}}\r)}\r)\r.
\l.+2\,B\,\tilde{{\cal M}}(x)\r]W(x)-2\,B\,x^3\,\w^2\,\tilde{\phi}(x)^2=0\ ,\\
\end{split}
\ee
\be
\begin{split}
\frac{x-2\,\tilde{{\cal M}}(x)}{x}\,\tilde{\phi}''(x)+\frac{2}{B\,x^2}\l[x\l(B-B\,\tilde{{\cal M}}(x) \r.\r. \\
\l.\l.-8\pi\,\tilde{f}_{a}^2\,x^2\l(1-\sqrt{1-2\,B+2\,B\,\cos\l(\frac{\tilde{\phi}(x)}{2\sqrt{\pi}\,
f_{a}}\r)}\r)\r)\r]\tilde{\phi}'(x)\\
+\frac{\w^2}{W(x)}\,\tilde{\phi}(x)-\frac{2\sqrt{\pi}\,\tilde{f}_{a}}{\sqrt{1-2\,B+2\,B\,\cos\l(\frac{
\tilde{\phi}(x)}{2\sqrt{\pi}\,f_{a}}\r)}}\,\sin 
\l(\frac{\tilde{\phi}(x)}{2\sqrt{\pi}\,f_{a}}\r)=0\ .
\end{split}
\ee

\bibliographystyle{JHEP}
\bibliography{References}

\end{document}